\documentclass[aps,,pre,twocolumn,showpacs,amsmath,amssymb,floatfix]{revtex4}
\usepackage{graphicx,color, bm} 
\usepackage{dcolumn} 

\newcommand{\be}{\begin{equation}}
\newcommand{\ee}{\end{equation}}

\begin{document}

\title{Forced MHD turbulence in three dimensions using Taylor-Green symmetries}

\author{G. Krstulovic$^1$, M.E. Brachet$^2$ and A. Pouquet$^{3,4}$}
\affiliation{
$^1$ \ Laboratoire   Lagrange, UMR7293, Universit\'e de Nice Sophia-Antipolis, CNRS, Observatoire de la C\^ote d'Azur, BP 4229, 06304 Nice Cedex 4, France\\
$^2$  \ Laboratoire de Physique Statistique de l'\'Ecole Normale Sup\'erieure, associ\'e au CNRS et aux Universit\'es ParisVI et VII, 24 Rue Lhomond, 75231 Paris, France. \\
$^3$Laboratory for Atmospheric and Space Sciences, and Department of Applied Mathematics, University of Colorado, Boulder, CO 80309, USA.  \\ 
 $^4$Computational and Information Systems Laboratory, NCAR,  Boulder CO 80307, USA.
}

\begin{abstract}
We examine the scaling laws  of MHD turbulence for three different types of forcing functions and imposing at all times the four-fold symmetries of the Taylor-Green (TG) vortex generalized to MHD; no uniform magnetic field is present and the magnetic Prandtl number is equal to unity.  We also include a forcing in the induction equation, and we take the three configurations studied in the decaying case in [E. Lee et al. Phys. Rev.E {\bf 81}, 016318 (2010)]. 
To that effect, we employ direct numerical simulations up to an equivalent resolution of $2048^3$ grid points.
We find that, similarly to the case when the forcing is absent,  different spectral indices for the total energy spectrum emerge, corresponding to either a Kolmogorov law,  an Iroshnikov-Kraichnan law that arises from the interactions of turbulent eddies and Alfv\'en waves, or to weak turbulence when the large-scale magnetic field is strong. We also examine the inertial range dynamics in terms of the ratios of kinetic to magnetic energy, and of  the turn-over time to the Alfv\'en time, and analyze the temporal variations of these quasi-equilibria.
\end{abstract}
\maketitle

\section{Introduction}\label{s:intro}
Turbulence is a common feature of  a variety of flows, from engineering to geophysics and astrophysics. It remains unsolved, due in part to a lack of statistical theory on how to deal with a very large number of modes interacting nonlinearly,
and competing with waves.
 At the moderate Reynolds numbers that are achievable today 
numerically
in three space dimensions on uniform grids of at most $4096^3$ points, one follows accurately the temporal  evolution of in excess of 64 billion modes, leading to the creation of myriads of vortex filaments. 
When coupling to a magnetic field, in the magnetohydrodynamic (MHD) approximation regime for velocities small compared to the speed of light, one observes current and vorticity sheets, that are found to roll-up for sufficiently high Reynolds numbers \cite{roll}. 

One question concerns the universality or not of the scaling laws of turbulent flows. There has been much debate concerning this point, in particular in the MHD community as well as when dealing with the dynamics of the atmosphere and the oceans: one way to phrase the question is to ask wether the presence of waves will affect the energy distribution among modes, inertial waves in the rotating case with solid body rotation, gravity waves for stratified turbulence, Alfv\'en waves in MHD, acoustic waves when the condition of incompressibility is removed, as is necessary in the interstellar medium where supersonic flows are routinely observed. The answer is unambiguous in the regime of weak turbulence when the ratio of characteristic times (wave period and eddy turnover time) is small; this small parameter allows for a natural closure to the statistical problem and constant-flux (as well as zero flux) solutions can be found in terms of power laws as a function of anisotropic wave numbers, the anisotropy arising from the imposition of an external agent (uniform rotation, gravity, or magnetic field), and to the anisotropic dispersion relations \cite{note}. But this weak-turbulence regime is non-uniform in scale, simply because the variation with scale of the wave period $ \tau_W$ and of the eddy turn-over time $\tau_{NL}$ are different; hence, there exists  a scale at which these two timescales are equal and the weak turbulence regime breaks down. For stratified flows, this is called the Ozmidov length scale, and for the rotating case, the Zeman scale.
Note that for MHD, the situation is different: for stratified and rotating flows, at scales smaller than the Ozmidov or Zeman scales, isotropy and a classical Kolmogorv scaling is likely to recover \cite{3072}, whereas it does  not in MHD. In fact, one could argue the opposite: isotropy can prevail at large scales where the effect of the large-scale magnetic field is purely local and its amplitude is comparable to that of the modes it is interacting with, whereas the anisotropic effect due to the imposed large-scale magnetic field is strong at small scale unless reconnection processes are numerous and random enough that isotropy again is recovered. This point is still in debate.

In MHD, another hypothesis has been put forward to understand the dynamical exchanges in a phenomenological way, that of an equality between the two characteristic time scales, an equality that would hold throughout the inertial range \cite{GS}. This hypothesis  leads to a Kolmogorov spectrum $E(k_{\perp}) \sim k_{\perp}^{-5/3}$ (hereafter Kp41), expressed in terms of $k_{\perp}$, where the direction refers to that of the external agent, here a uniform 
magnetic field. It was found in \cite{mangeney} that the same hypothesis  can also lead to an Iroshnikov-Kraichnan spectrum (IK hereafter) or a weak turbulence spectrum, 
$E(k_{\perp}, k_{\parallel})\sim k_{\perp}^{-2}f(k_{\parallel})$ (WT hereafter), on the simple basis that $\tau_W(k)/\tau_{NL}(k)=r_{\tau}(k)$ can be constant but not necessarily equal to unity: the different regimes appear in that light as emerging from a different rate at which energy is exchanged between its kinetic and magnetic modes. All these spectra have been observed in direct numerical simulations (DNS) of three-dimensional (3D) MHD turbulence. In one particular 
case,  identical velocity fields are used as initial conditions, with comparable invariants (total energy $E_{\rm tot}=0.25$ with $E_v (t=0)=E_b(t=0)$ where $E_{v,b}$ are the 
kinetic and magnetic energy respectively, with strictly zero magnetic helicity and negligible cross-correlation between the velocity and the magnetic field) \cite{lee2}.
It is the purpose of this paper to pursue the work done in \cite{lee2}, extending it to the statistically steady case in the presence of forcing.

In the next section, we write equations, initial conditions and forcing; the results are given for a high-resolution run in \S \ref {s:2048} and in \S \ref{s_1024} we compare the evolution of three 
magnetic configurations,
 at resolutions of $1024^3$ grid points. Finally, \S \ref{s:conclu} is the conclusion.

\section{The equations and the numerical set-up}\label{ss:eqs}

The MHD equations for an incompressible fluid with $\bf {v}$ and $\bf {b}$ respectively the velocity and magnetic fields in Alfv\'enic units are:
\begin{eqnarray}
&& \frac{\partial {\bf v}}{\partial t} + {\bf v} \cdot \nabla {\bf v} = 
    -\frac{1}{\rho_0} \nabla {\cal P} + {\bf j} \times {\bf b} + \nu \nabla^2  {\bf v} + {\bf F}_V , 
\label{eq:MHDv} \\
&& \frac{\partial {\bf b}}{\partial t} = \nabla \times ( {\bf v} \times
    {\bf b}) +\eta \nabla^2 {\bf b} + {\bf F}_M \ ; 
\label{eq:MHDb}
\end{eqnarray}
$\rho_0=1$ is the (uniform) density, and ${\bf b}$ is  dimensionally a velocity as well, the Alfv\'en velocity; ${\cal P}$ is the total pressure,
${\bf \nabla} \cdot {\bf v} = \nabla \cdot {\bf b} = 0$, and $\nu$ and $\eta$ 
are respectively the kinematic viscosity and magnetic diffusivity;  we take $\nu=\eta$ (unit magnetic Prandtl number). Finally, ${\bf F_V, F_M}$ are forcing terms introduced both in the momentum and in the induction equations. In principle, above a critical Reynolds number $R_M^C$, a dynamo mechanism sets in whereby sufficient magnetic excitation is produced at all scales. For the Taylor-Green flow defined below it was shown in reference \cite{giorgio_11} that $R_M^C$ depends very strongly on the imposed symmetries. In addition, when imposing all symmetries at all times, it was shown in reference \cite{nore} that $R_M^C$ is very high (of the order of 1000). 
In this work, like in reference \cite{lee2}, we focus of the fully symmetric problem in order to maximize the available 
Reynolds number, and thus the maximum
resolution (see discussion following Eq. (\ref{descompfourier}) below).
Hence, we chose to force the induction equation as a way to mimic the dynamo itself. We note that, by simply breaking the symmetry of the initial conditions and using a general code, this critical parameter is lowered by more than one order of magnitude, but with a substantially costlier computation, 
by a factor of 32 \cite{nore}. The forcing in the induction equation, $F_M$, is not a common choice. It is included in order to compensate for the fact that, in the presence of symmetries, the generation of a magnetic field by fluid turbulence (or dynamo effect) occurs above a threshold in magnetic Reynolds number of $R_b \approx 1000$, and is slow in this vicinity of $R_b$. Another justification for $F_M\not= 0$ comes from the dynamics of 
the Solar Wind \cite{perez_10, bigot_11}, with Alfv\'en wave forcing stemming from coronal mass ejections.

The energy $E_{\rm tot}$, the cross helicity $H_C$ and the magnetic helicity $H_M$ are defined as
\begin{eqnarray}
E_{\rm tot}=E_v+E_b&=&\left<v^2+b^2\right>/2 \\
H_C=\left<{\bf v} \cdot {\bf b}\right> &,& H_M=\left<{\bf A} \cdot {\bf b}\right> 
\end{eqnarray}
where  ${\bf b} = \nabla \times {\bf A}$ and ${\bf A}$ is the magnetic potential. 
In the ideal case
($\nu=\eta=0$) and without forcing (${\bf F}_V={\bf F}_M=0$)
note that these quantities ($E_{tot}, H_C$ and $H_M$) are all conserved.
Relative helicities measure the relative alignment of vectors, independently of their amplitudes, 
$\rho_V= \cos{[{\bf v}, {\bf \omega}] }\ , \rho_C= \cos{[\bf v}, {\bf b}] \ , \ \ \rho_M = \cos{[\bf A}, {\bf b}] $, with ${\bf \omega} = \nabla \times {\bf u}$ the vorticity (the kinetic helicity $H_V=\langle {\bf u} \cdot {\bf \omega} \rangle$ is an invariant when ${\bf b}\equiv 0$).

Considering a flow which is $2 \pi$-periodic in all spatial dimensions, the kinematic Reynolds number ${\rm Re}$ and the magnetic Reynolds number  ${\rm R_m}$ are defined as
\begin{equation}
{\rm Re}=\frac{L v_{\rm rms}}{\nu},\hspace{.5cm} {\rm R_m}=\frac{L v_{\rm rms}}{\eta}\label{Eq:defReynolds}
\end{equation}
where the root-mean square velocity is $v_{\rm rms}=\sqrt{2 E_v /3}$ and the characteristic length $L$ is defined by 
\begin{equation}
L=2\pi\frac{\sum_{k} k^{-1}E_v (k,t)}{\sum_{k} E_v (k,t)\,dk},
\end{equation}
where  the kinetic energy spectrum $E_v (k,t)$ (such that $E_v (t)=\sum_{k}E_v (k,t)$) is obtained by summing
${\frac1 2}|{\bf \hat u}({\bf k'},t)|^2 \,$  on the spherical shells $k-1/2\leq |{\bf k'}|<k+1/2$
(${\bf \hat u}({\bf k})$ is
the Fourier transform of the velocity).
Analogously, the magnetic energy spectrum is denoted by $E_b (k,t)$ and verifies $E_b (t)=\sum_{k} E_b (k,t)$.

We now turn to the definition of the external driving volumic forces ${\bf F}_{V,M}$ in \eqref{eq:MHDv}-\eqref{eq:MHDb} which balance the total energy dissipation and allow to reach a statistically stationary state. 
Following reference \cite{nore}, we force the system by setting in \eqref{eq:MHDv} 
\begin{equation}\label{Eq:defForcev}
{\bf F_V}=f_v {\bf v^{\rm TG}},
\end{equation}
where ${\bf v^{\rm TG}}$ is the Taylor-Green vortex \cite{TG} given by
\begin{equation} {\bf v^{\rm TG}}=(\sin(x) \cos(y) \cos(z),-\cos(x) \sin(y)\cos(z), 0), \label{eqn:simple_vTG}\end{equation}
and $f_v$ is always set to the value $1/16$.
The force ${\bf F_M}$ is determined in a similar way but using instead of the TG velocity mode, the three magnetic field modes studied in the decaying MHD runs of reference \cite{lee2}. 
These three modes were:
 \begin{equation}
{\bf b^I}=b_0^I \begin{pmatrix}
 \cos x\sin y\sin z \\
 \sin x\cos y\sin z  \\
-2 \sin x\sin y\cos z
 \end{pmatrix},\label{eqn:btg_I}
\end{equation}
\begin{equation}
{\bf b^A}=b_0^A \begin{pmatrix}
\cos2x\sin2y\sin2z \\
-\sin2x\cos2y\sin2z  \\
0
 \end{pmatrix},
\end{equation}
\begin{equation}
{\bf b^C}=b_0^C\begin{pmatrix}
\sin2x\cos2y\cos2z \\
\cos2x\sin2y\cos2z  \\
-2 \cos2x\cos2y\sin2z
 \end{pmatrix}. \label{eqn:btg_C}
\end{equation}
The labels  $I$, $A$ and $C$ stand respectively for insulating, alternate insulating and conducting boundary conditions for the current when considering its orientation with respect to the wall of the so-called fundamental box (see \cite{brachet_TG}). The coefficients $b_0^I $, $b_0^A$ and $b_0^C$ are such that, for all cases, $E_b=1/16$. Thus, all  computations have equal initial kinetic and magnetic energy at t=0, with $E_{\rm tot}=E_v +E_b=1/8$.

Like in Eq. \eqref{Eq:defForcev}, the amplitudes of the forcing are chosen as
\begin{equation}\label{Eq:defForcem}
{\bf F_M}=f_b {\bf b^{\rm I,A,C}},
\end{equation}
and the values of $f_b$ are given below in Table \ref{RUNS}.

We shall also examine the behavior of the spectral ratios of time-scales $R_{\tau}$ and of modal energies $R_E$ defined respectively as
\begin{equation}
R_{\tau}(k)=\tau_{NL}(k)/\tau_A(k) , \ \ \ R_{E}(k)=E_b(k)/E_v (k) \ , 
\label{eq:tau} \end{equation}
with $\tau_{NL}(k)=[k\hat u(k)]^{-1}$ and $\tau_A(k)=[kB_0]^{-1}$, $B_0$ being defined here as the magnetic field in the gravest mode (the first non-zero mode).

Because of the symmetries of the TG vortex extended to MHD,
 all fields can be represented in Fourier space as:
\begin{eqnarray}\label{descompfourier}
  v_x(r,t) &=&\! \!\!\!\! \sum\limits_{m,n,p=0}^\infty \!\!\!\! u_x(m,n,p,t)\sin{mx}\cos{ny}\cos{pz} , \\
  v_y(r,t) &=&\! \!\!\! \! \sum\limits_{m,n,p=0}^\infty \!\!\!  u_y(m,n,p,t)\cos{mx}\sin{ny}\cos{pz}  ,  \\
  v_z(r,t) &=&\! \!\!\!\!  \sum\limits_{m,n,p=0}^\infty \!\!\!  u_z(m,n,p,t)\cos{mx}\cos{ny}\sin{pz}  ,  
\end{eqnarray}
where $\textbf{u}(m,n,p,t)$ is equal to zero unless the three integers $m,n,p$ are either all even or all odd. Thus, in spectral space, the following symmetry is fulfilled:
\begin{eqnarray}
  u_x(m,n,p,t) &=& (-1)^ru_y(n,m,p,t)\label{SymetTGSpec1} \ ,  \\
  u_z(m,n,p,t) &=& (-1)^ru_z(n,m,p,t)\label{SymetTGSpec2} \ , 
\end{eqnarray}
with $r=1$ if $m,n,p$ are all odd, and $r=2$ if $m,n,p$ are all even.

Relations (\ref{SymetTGSpec1}) and (\ref{SymetTGSpec2}) allow one to only compute 
$v_x$ and $v_z$. Moreover, the decomposition (\ref{descompfourier}) on either even or odd integers  leads to a gain of a factor of 32
in memory and CPU time compared with the general case of Fourier transforms with the same scale separation, or $k_{max}/k_{min}$, with respectively $k_{min}=1$ for a box of length $2\pi$ and $k_{max}=N/3$ with $N$ the number of grid points per dimension, using a standard 2/3 de-aliasing rule.
The code is pseudo-spectral, with a fourth-order Runge-Kutta temporal scheme and with periodic boundary conditions. All  previous symmetry relations are implemented to speed up calculations. The code is parallelized up to $\sim 98,000$ processors on grids of up to $8196^3$ points, using a hybrid (MPI-Open-MP) algorithm  which becomes advantageous at high resolution \cite{hybrid}. Grids used in this work have the equivalent resolution of $1024^3$ and $2048^3$ points.
Run parameters are summarized in Table \ref{RUNS}. 
\begin{table}
\begin{tabular}{| c || c | c | c | c | c | c |c|}  \hline   
  RUN & $N$         & $T_f$   & $dt$ 				& $\nu=\eta$  			& $f_b$ 				&$Re$   	        	  \\ 
  \hline
  C1     &  $2048$ & $22.5$ & $3.125\times 10^{-4}$ & $6.25\times 10^{-5}$	& $6.25\times 10^{-2}$     & $8700$         	  \\
  C2     &  $1024$ & $68$    & $6.25\times 10^{-4}$   & $1.25\times 10^{-4}$      & $6.25\times 10^{-2}$	& $4470$         	\\
  \hline
  I       &  $1024$ & $52$    & $6.25\times 10^{-4}$  	 & $6.25\times 10^{-5}$    & $10^{-4}$ 			& $1360$          \\
  \hline
  A     &  $1024$ & $52$    & $6.25\times 10^{-4}$ 	  & $2.5\times 10^{-4}$	& $6.25\times 10^{-2}$    				& $2270$      \\
    \hline  \end{tabular}   \caption{List of runs and parameters. Resolution $N$,  final time $T_f$ and time step $dt$, viscosity and magnetic diffusivity $\nu=\eta$, forcing parameter $f_b$  and Reynolds number $Re$ (see Eqs. (\ref{eq:MHDv}-\ref{eq:MHDb},\ref{Eq:defReynolds} and \ref{Eq:defForcem})).
     \label{RUNS}} 
      \end{table}
To correctly resolve the MHD equations spectrally, a fast decay at large $k$ (faster than algebraic) of the energy spectrum is required. This condition (called spectral convergence) is quantitatively determined by fitting the exponential decay of the energy spectra by a law of the form $C  e^{- 2 \delta k}$ that amounts to a simple Lin-Log linear regression. The value of  $\delta k_{\rm max}$ furnishes a measure of spectral convergence. We obtain values of $\delta k_{\rm max}$ of $5.4$, $3.37$, $2.3$ and $5.2$  for the runs C1, C2, I and A respectively, showing that all simulations all well resolved.

We would like to remark that if symmetries are not enforced, due to round noise a symmetry breaking can take place, as studied in \cite{Alexakis_Dallas_SymBreak}. Therefore, during the statistical stationary regime reached after a very long time, systems with and without imposed symmetries are not equivalent. The advantage of imposing the TG symmetries is not purely numerical, it also provide a way to mimic more realistic boundary condition for both, velocity and magnetic field (for a long discussion see for instance \cite{giorgio_11}).

\section{Results for the C runs} \label{s:2048}

\begin{figure} [h!]
\includegraphics[width=0.95\columnwidth]{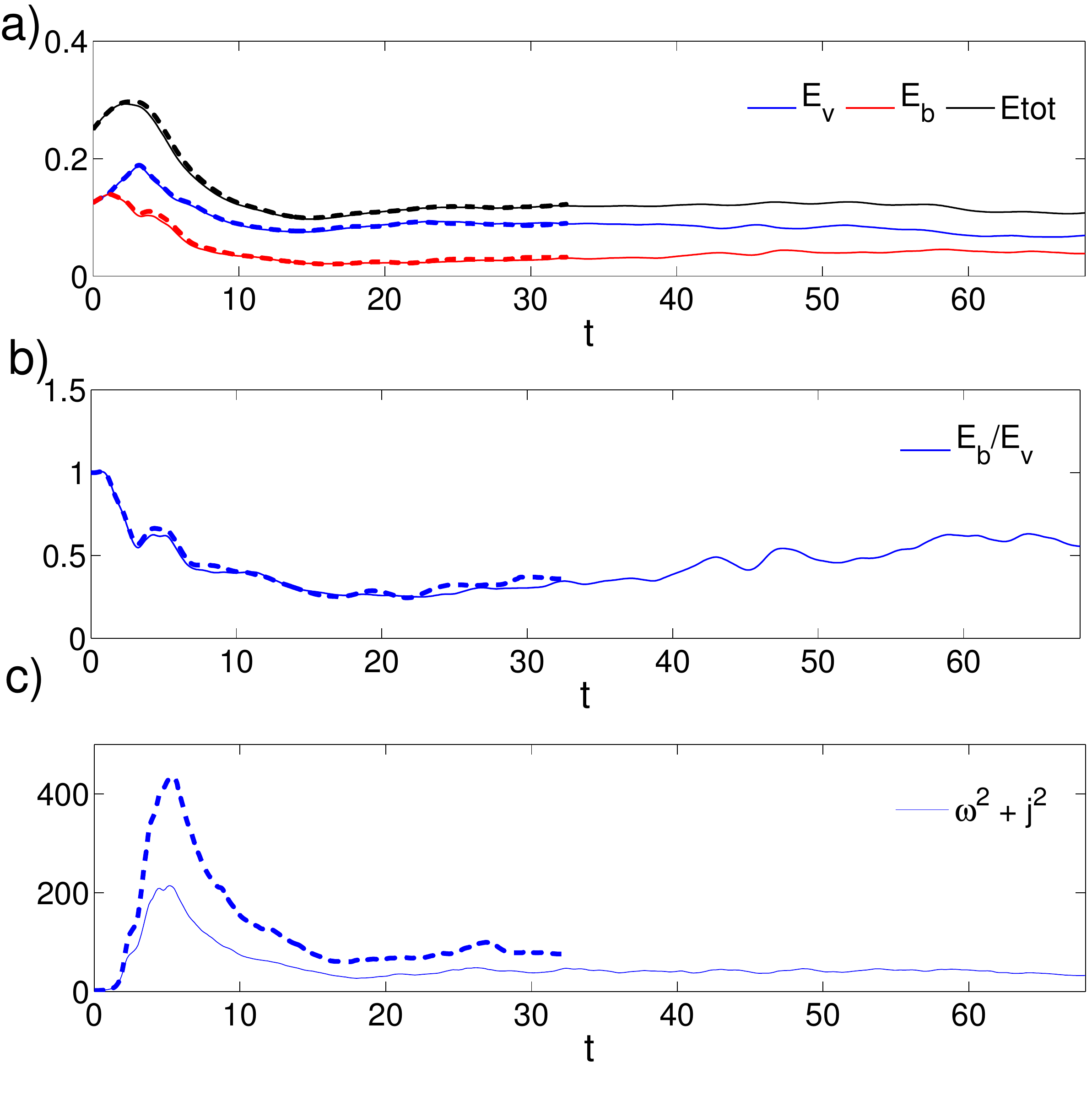}
\caption{ (Color online) 
{\it  a:} Temporal evolution of kinetic (blue), magnetic (red) and total (black) energy for Runs C1 and C2 (see Table \ref{RUNS});
{\it  b:} Magnetic to kinetic energy ratio as a function of time for the same runs; this ratio seems to saturate at a value $\sim 0.55$. 
{\it  c:} Temporal evolution of $\omega^2+j^2$ where $\omega=\nabla\times {\bf u}$ is the vorticity and $j=\nabla\times {\bf b}$ the magnetic current. 
In all figures, longer temporal evolutions up to $t\sim 65$, are given for the run on a grid of $1024^3$ points.
}  
\label{f_temp_2C} 
 \end{figure}

The first run on which we report is the one at the highest Reynolds number (and the highest resolution).  We give in Fig. \ref{f_temp_2C}.a, top, the temporal evolution of the kinetic, magnetic and total energy. 
 The thick lines are for Run C1 and the dashed lines for run C2 at lower resolution (see Table \ref{RUNS}). The ratio of magnetic to kinetic energy $R_E$ is given in Fig. \ref{f_temp_2C}.b. The total energy  saturates for times larger than $\sim 20$, at $\sim 0.12$, and the global energetic ratio is between $R_E\sim 0.3$ and $R_E\sim 0.6$, with a tendency toward growth. Figure \ref{f_temp_2C}.b displays the temporal evolution of $\omega^2+j^2$ where $\omega=\nabla\times {\bf u}$ is the vorticity and $j=\nabla\times {\bf b}$ the magnetic current. This quantity also saturates showing that small scales also converged to a statistically stationary state.

\begin{figure}[h!] 
 \includegraphics[width=0.95\columnwidth]{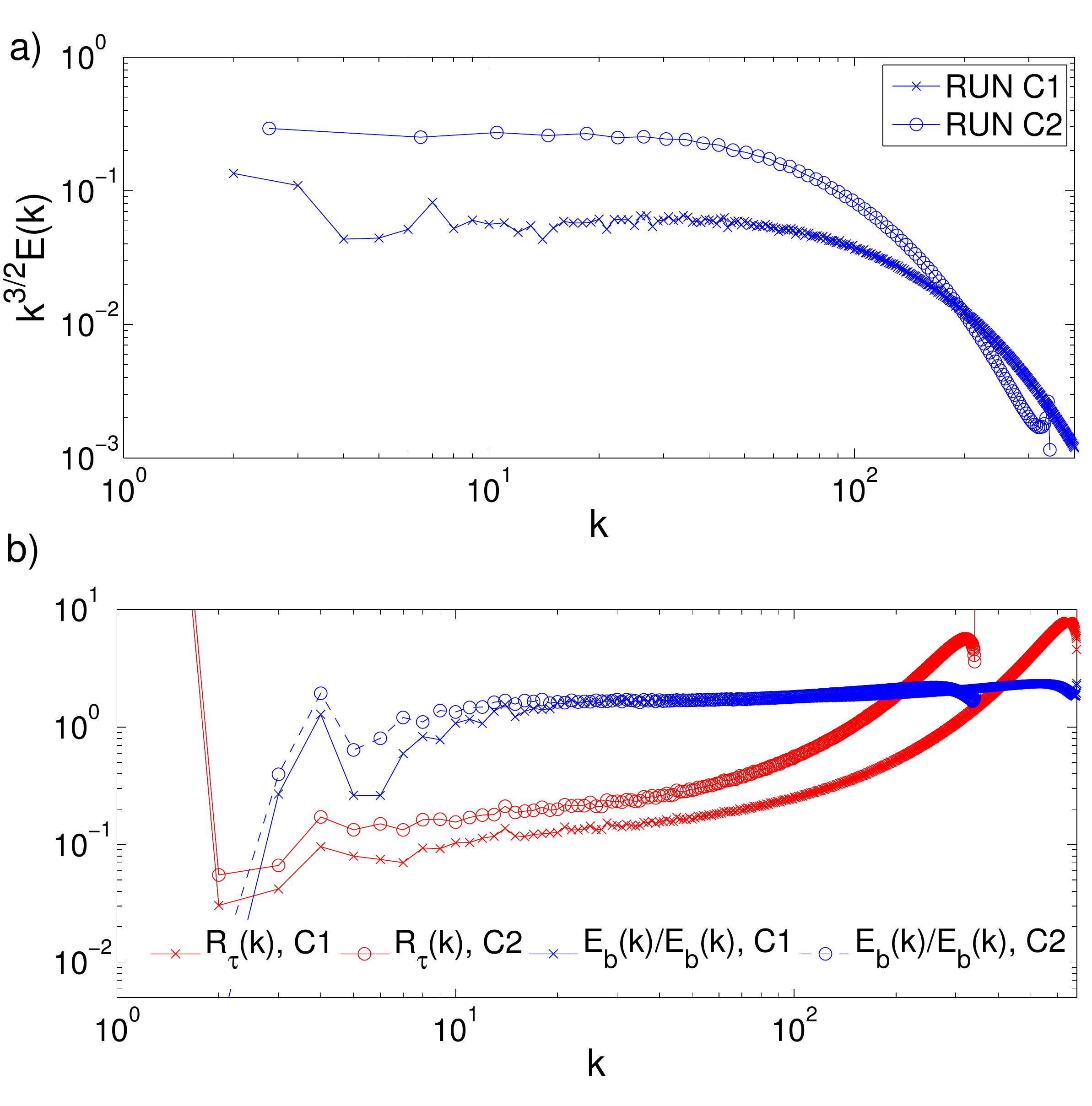}
\caption{ (Color online) 
 For runs $C1$ and $C2$ (see Table \ref{RUNS}), 
a) 
 $k^{3/2}$-compensated total energy spectrum temporally averaged in the interval $t=17-22.5$ for run C1 and $t=20-68$ for run C2.
b) Spectral ratios of energy $R_E(k)$ (blue) and time-scales $R_{\tau}(k)$ (red), as defined in Eq. (\ref{eq:tau}).
Note the constancy of the former in the inertial range.
} \label{f_spec_2C}  
\end{figure}

We performed a temporal average of the total energy spectrum from $t=17.4$ to $t=22.5$ for runs C1 (red, solid line) and C2 (blue, dashed line) which we present in Fig. \ref{f_spec_2C}.a, compensated by $k^{+3/2}$.
As in the decay case presented in \cite{lee2}, the best fit is for $E_{\rm tot}(k) \sim k^{-3/2}$, i.e. an Iroshnikov-Kraichnan law (See table \ref{RUNS_2} for details).
We give further below the temporal evolution of the instantaneous value of the spectral index  in Fig.\ref{Fig:Compare}.b (bottom), when comparing it for several runs (see next Section).

\begin{table}
\begin{tabular}{| c ||  c | c | c |}  \hline   
  RUN          & $t_i - t_f$   & $k_i -k_f$ 				& {\rm exponent}    \\ 
  \hline
  C1          & $20-32$   & $5-50$ 				& $1.448\pm 0.025$  \\ 
   C2         & $30-60$   & $5-30$ 				& $1.516\pm0.028$   \\ 
    \hline 
     I        & $35-52$   & $10-38$ 				&$1.6334\pm0.032$   \\ 
  \hline  
       A         & $30-52$   & $5-35$ 				&$ 1.977\pm0.051$    \\ 
      \hline  
  \end{tabular} 
    \caption{Temporal and spectral ranges used for the power-laws fits.
     \label{RUNS_2}} 
      \end{table}

In Fig. \ref{f_spec_2C}.b is shown the ratio of the spectra of the turn-over time to the Alfv\'en time, $R_{\tau}$, defined in Eq. (\ref{eq:tau}) and averaged in the same temporal interval as the energy spectrum.  As can be seen from Fig.\ref{f_spec_2C}.b, there is a systematic increase of this ratio in the inertial range, contrary to the hypothesis of critical balance advocated in \cite{GS} (see also \cite{mangeney, nazar}). Rather, it is the 
  magnetic to kinetic spectra ratio which remains  remarkably constant throughout the inertial range, as displayed in Fig.\ref{f_spec_2C}.b, with as often, a slight excess of magnetic energy, except at the gravest mode that dominates the global energetics; this confirms the earlier findings of the decay case \cite{lee2}, as well as those in numerous other numerical simulations (see e.g. \cite{julia}).

When examining the behavior of the C-flow at lower resolution on a grid of $1024^3$ points,  we observe that the results are in agreement with these conclusions; the lower resolution simply allows us to compute for longer times, leading to a better temporal averaging. Nevertheless, there may be a trend toward the energetic ratio to increase at later times (see Fig. \ref{f_temp_2C}).

\section{Comparative results for the three forcing functions } \label{s_1024}

The most
 striking result of the computations performed in the decaying case presented in \cite{lee2} is that different initial conditions 
 for the magnetic field only, but with the same global invariants,
   led to different spectral inertial indices. Will the same occur in the presence of forcing? This is what we are now investigating. We thus 
 address now the question of the scaling of the two other configurations studied in \cite{lee2} in the decaying case, namely the so-called A- and I- magnetic configurations.

 \subsection{The A run}
\begin{figure}[h!]
\includegraphics[width=0.95\columnwidth]{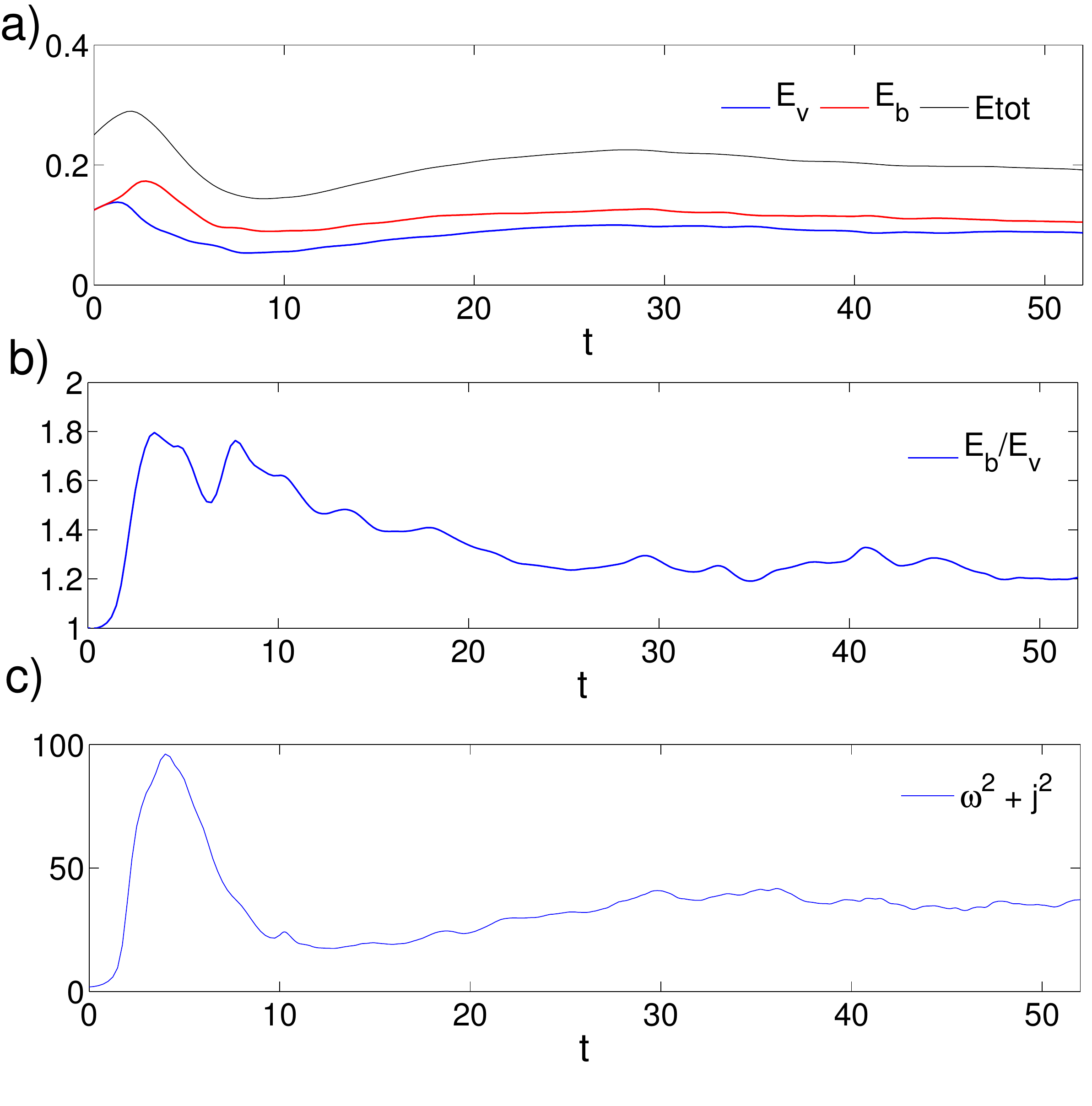}
\caption{(Color online) 
a)Temporal evolution of kinetic (blue), magnetic (red) and total (black) energy for Run A (see Table \ref{RUNS}). 
b) Temporal evolution of the ratio $E_b /E_v $ for the same run.
Equipartition is almost reached in that run, with a slight excess of magnetic energy.
{\it  c:} Temporal evolution of $\omega^2+j^2$ where $\omega=\nabla\times {\bf u}$ is the vorticity and $j=\nabla\times {\bf b}$ the magnetic current.
}
 \label{f_temp_1A} 
  \end{figure}

\begin{figure}[h!] \begin{center}
\includegraphics[width=0.98\columnwidth]{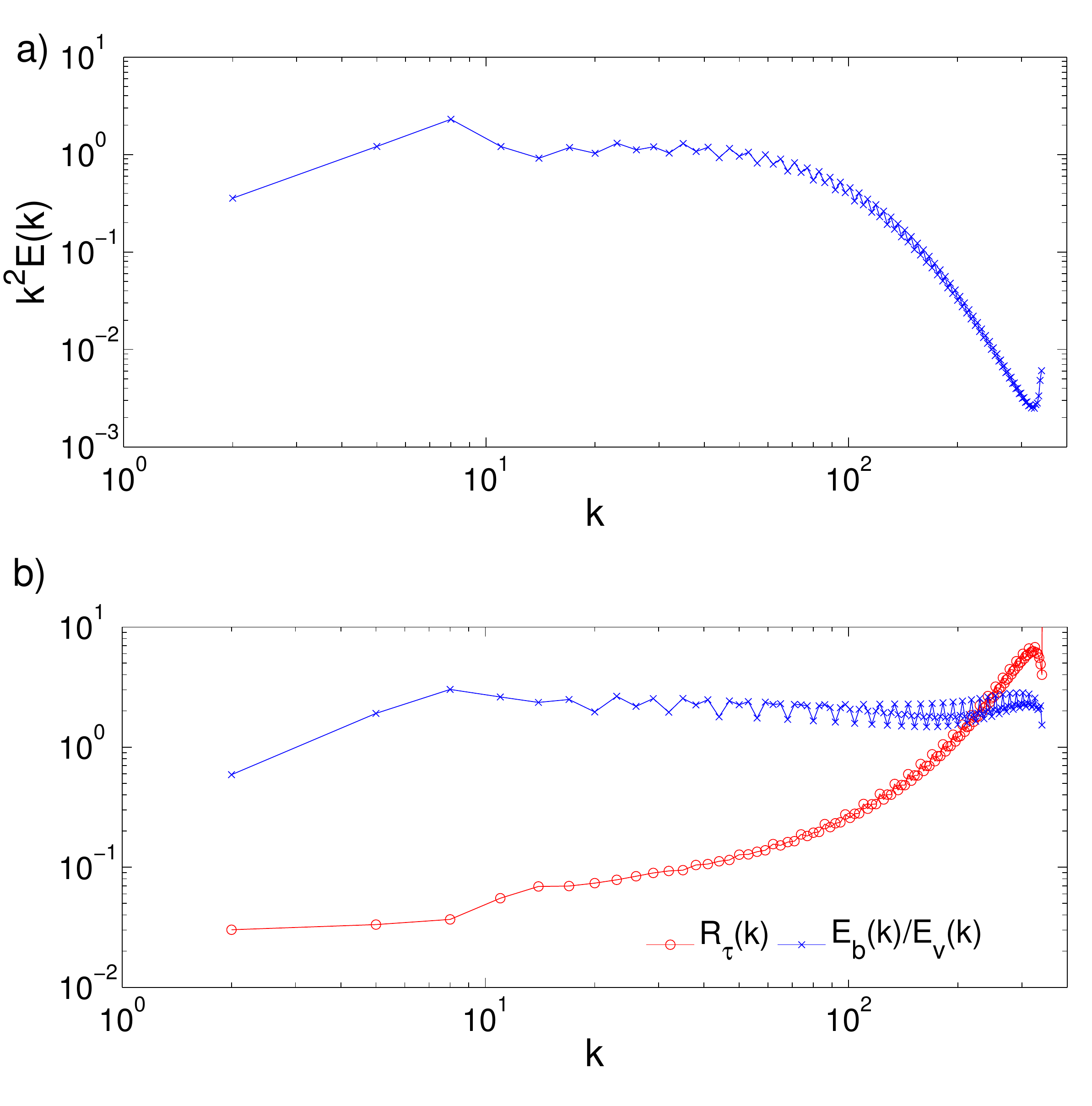}
\caption{(Color online) 
 Run $A$ (see Table \ref{RUNS}); all spectra are averaged in time in the interval $t=25-52$ and computed by summing on spectral shells of width $\Delta k=3$ to reduce oscillations.  a) 
 $k^{2}$-compensated total energy spectrum.
 b) Ratios of energy spectra (red), 
 again constant in the inertial range, and ratio of time-scales (blue).
} \label{Fig:A} \end{center} \end{figure}

 The forcing with the A configuration leads the system to reach a statistically stationary state, both globally for $E_{\rm tot}$ (shown in Fig. \ref{f_temp_1A}.a) and in its kinetic to magnetic energy ratio (as displayed in Fig. \ref{f_temp_1A}.b), with a value of that ratio slightly above unity, as often observed in the Solar Wind.
Figure \ref{f_temp_1A}.b also displays the temporal evolution of $\omega^2+j^2$, a clear stationary state is observed for $t\gtrsim30$.
The total energy spectrum for this run is rather steep,
with the best fit corresponding to a
$k^{-2}$ law (see Fig. \ref{Fig:A}.a), i.e. a law corresponding to weak turbulence (See table \ref{RUNS_2} for details). This is in contrast to \cite{lee2}: in the decaying case, this flow had a spectral index close to $-5/3$. Note that the wave-turbulence behavior in the present case is consistent with the ratio of energy spectra and  the ratio of time scales displayed on Fig..\ref{Fig:A}.b. 
Note also that this power law can be attributed to the presence of a (quasi)-discontinuity in the magnetic field, as recently found for the decaying case for the I-flow, in the absence of imposed symmetries \cite{Alexakis_Dallas_Km2}.

In Fig. \ref{Fig:Compare}.b below the temporal evolution of the spectral indices of all the runs are compared. 
Note that the spectral index of the total energy spectrum of the A-flow varies substantially over time. 
At this point, it is difficult to decide which power law is best followed. Initial behaviour seems consistent with the unforced $-5/3$ value although the inertial law is steepest at later times, and is thus more in favor of a weak turbulence spectrum.
However note that steep structures, such as sharp and isolated current and vorticity sheets, can also lead to a ``shock'' like spectrum. In this context, see the visualizations presented below in Fig. \ref{Fig:Viz} at the end of the present section.

 \subsection{The I runs}

\begin{figure}[h!] \begin{center}
\includegraphics[width=0.95\columnwidth]{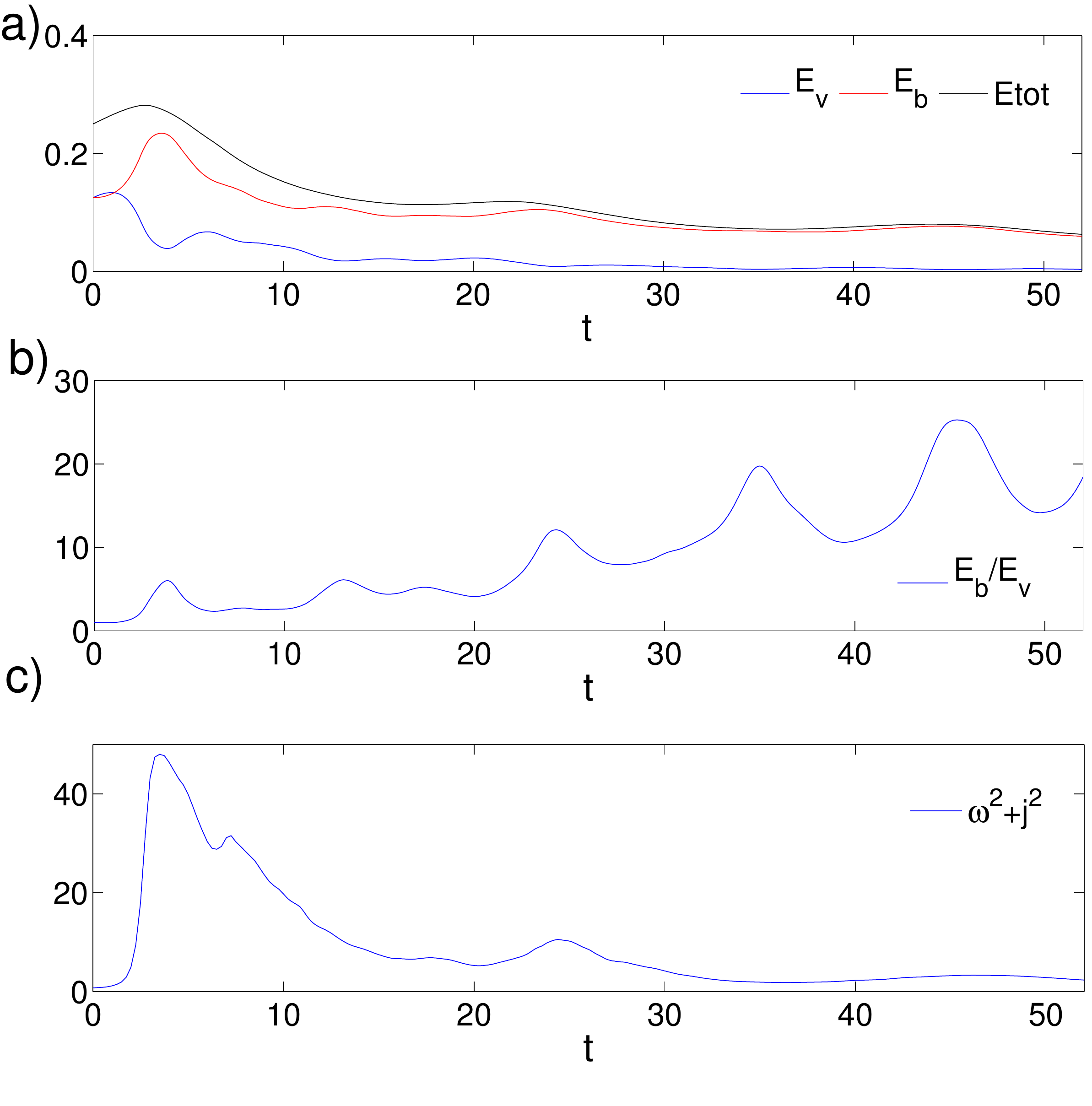}
\caption{ (Color online) 
a)Temporal evolution of kinetic (blue), magnetic (red )and total (black) energy for Run $I$ (see Table \ref{RUNS}). 
b) Temporal evolution of the ratio $E_b /E_v $ for the same run.
{\it  c:} Temporal evolution of $\omega^2+j^2$ where $\omega=\nabla\times {\bf u}$ is the vorticity and $j=\nabla\times {\bf b}$ the magnetic current. 
} \label{Fig:runIa} \end{center} \end{figure}

Let us now examine the dynamics of the magnetic I configuration (see definition in (\ref{eqn:btg_I})).
The temporal evolutions of the energies is displayed in Fig.\ref{Fig:runIa}.a.
The total energy and  $\omega^2+j^2$ seem to reach a quasi steady state although 
an increase of the ratio of magnetic to kinetic energy is observed in Fig.\ref{Fig:runIa}.
Thus, a stationary state is not reached for that ratio, with marked global oscillations.

\begin{figure}[h!] \begin{center}
\includegraphics[width=0.95\columnwidth]{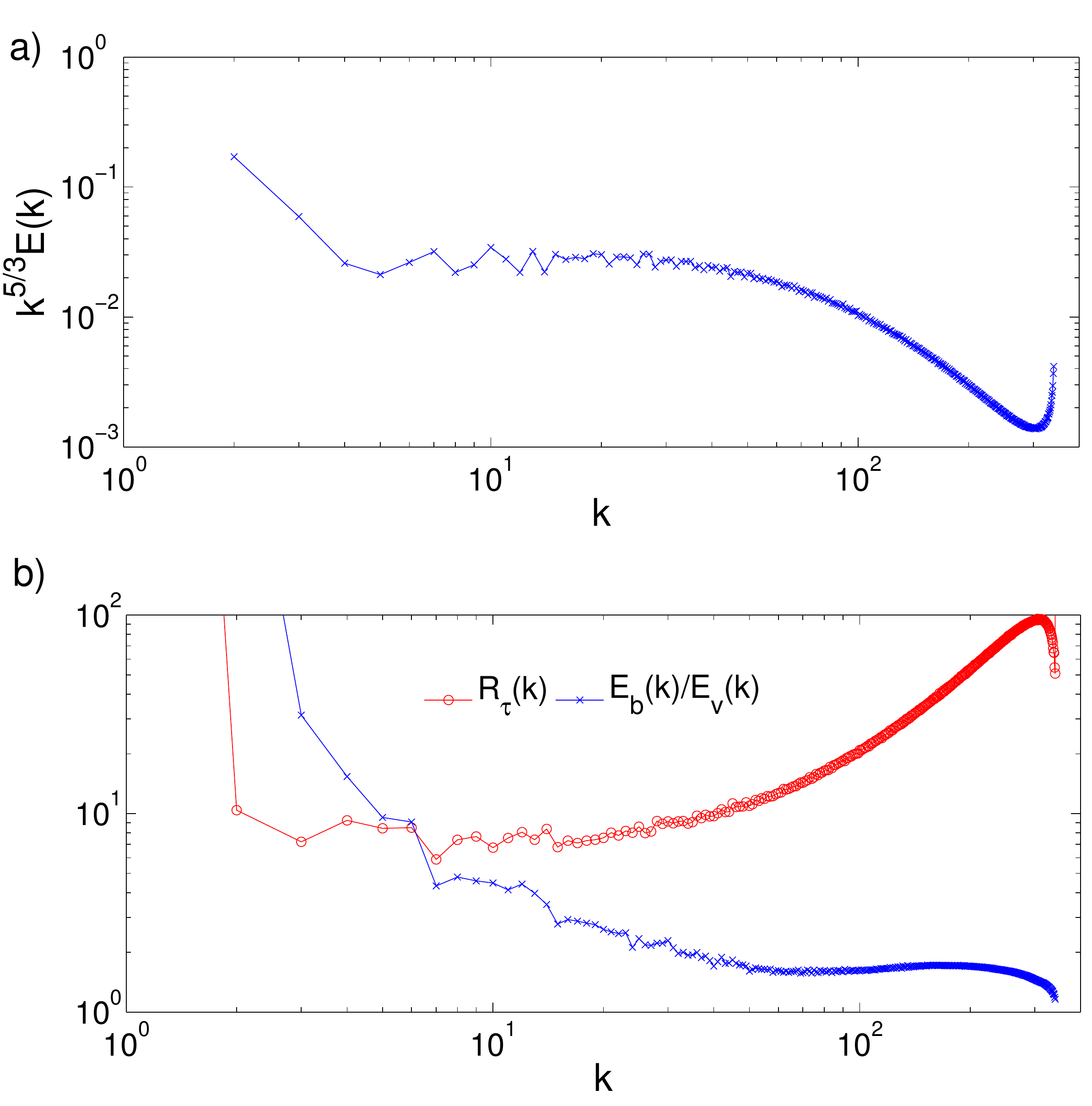}
\caption{ (Color online)  
 Run $I$ (see Table \ref{RUNS}); all spectra are averaged in time in the interval $t=30-52$ , 
a) 
 $k^{5/3}$-compensated total energy spectrum.
 b) Ratios of energy spectra (blue) and time-scales (red).
 } \label{Fig:runIb} \end{center} \end{figure}

\begin{figure}[h!] \begin{center}
\includegraphics[width=0.95\columnwidth]{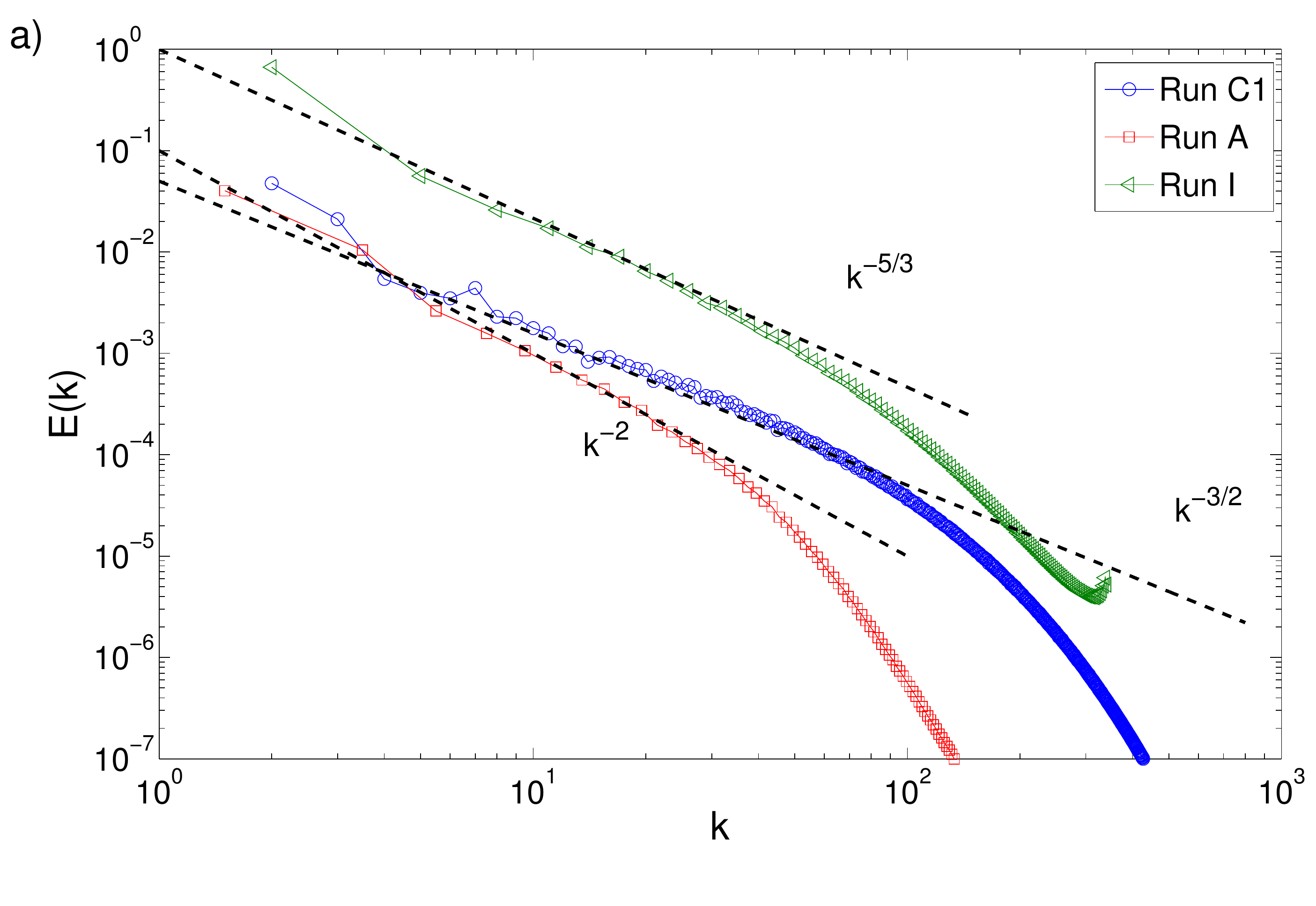}
\includegraphics[width=0.95\columnwidth]{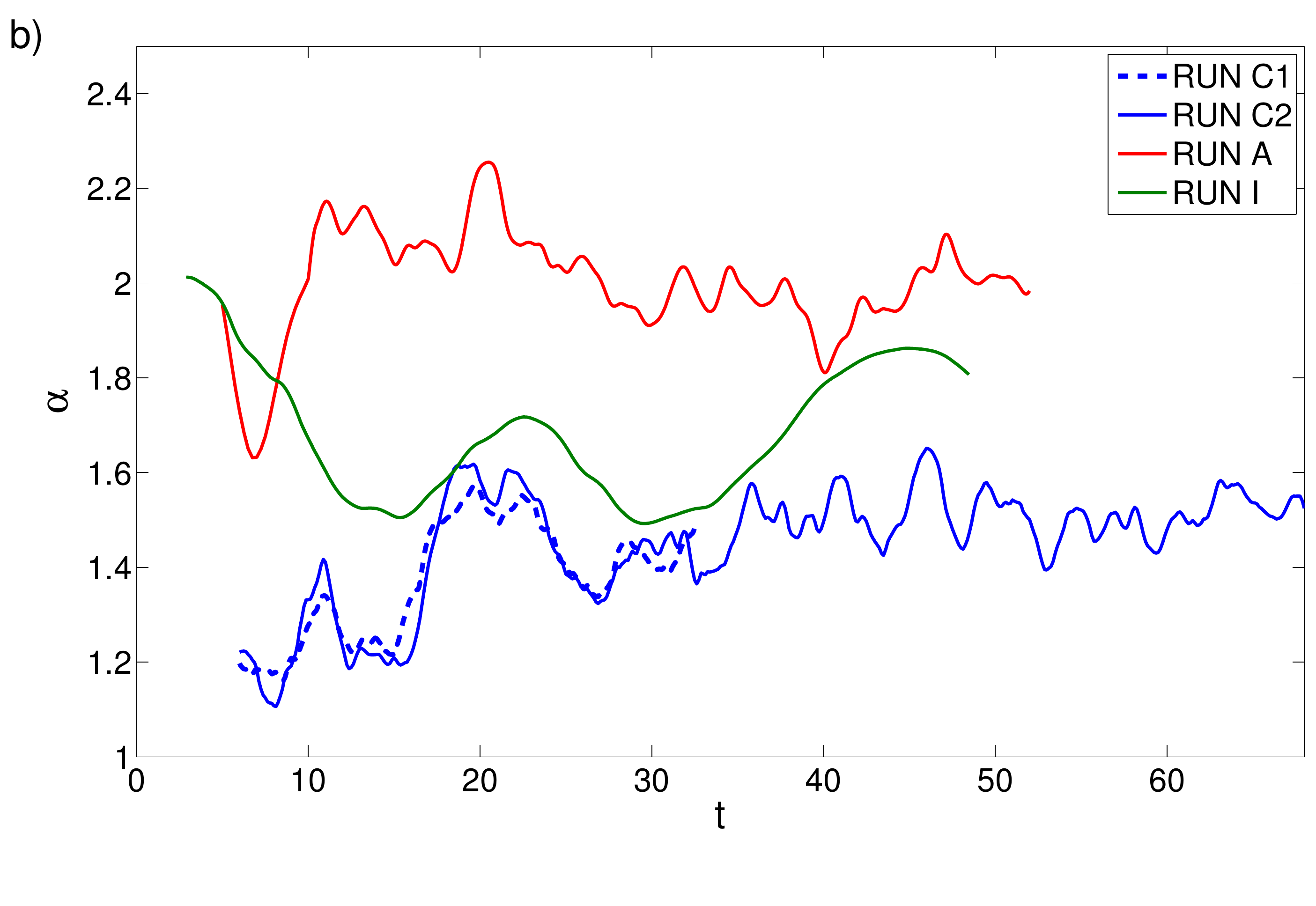}
\caption{ (Color online) a) Total energy spectrum for the runs $C2$, $A$ and $I$. Black dashed lines display the different fits.
b) Temporal evolution of spectral indices for all runs (see inset).
Note that the differentiation between the three magnetic configurations is clear at all times, even if the exact value of the indices vary with time.
} \label{Fig:Compare} \end{center} \end{figure}

The total energy spectrum (compensated by $k^{5/3}$) and the time-scales ratio and energy spectra ratio are shown in Fig.\ref{Fig:runIb}.
A correlation between the oscillatory growth of the  ratio of magnetic to kinetic energy displayed in Fig.\ref{Fig:runIa}.b and the spectral index of run I displayed as the middle curve in Fig.\ref{Fig:Compare}.b is apparent. Indeed
upward trends in the spectral index, moving towards weak turbulence, correspond to growth of magnetic energy at the expense of the kinetic energy. 
Observe in Fig.\ref{Fig:runIb}.b that magnetic energy dominates over the kinetic one and at the same time the ratio of time scales is almost constant in the inertial regime.

As already observed in \cite{lee2} in the decaying case, it may be that spectral indices vary with time, as the ratio of kinetic to magnetic energy varies as well. Also, it is quite difficult to distinguish between spectral indices that are quite close and this flow seems more undecided than for the other two flows we study in this paper.

To summarize, in Fig.\ref{Fig:Compare}.a we present the (uncompensated) spectra for the three runs (averaged over time).
We also show in Fig. \ref{Fig:Compare}b (bottom) the temporal variation of spectral indices for the three runs. 
We would like to emphasize that the actual values of the exponents shown in Table \ref{RUNS_2}, that were obtained after a time average, are not as significant  as  is their overall behavior displayed  in Fig. \ref{Fig:Compare}b. Indeed, fluctuations are observed because the scale separation between the inertial and dissipative ranges is not 
 large enough  and   finite resolution effects step in. It is important to remark here that, after an initial  transient,
   the exponents have each a well defined behavior and that in particular they do not overlap at any time for the three runs, revealing the non-universal character of MHD turbulence.

\begin{figure*}
\includegraphics[width=0.75\columnwidth]{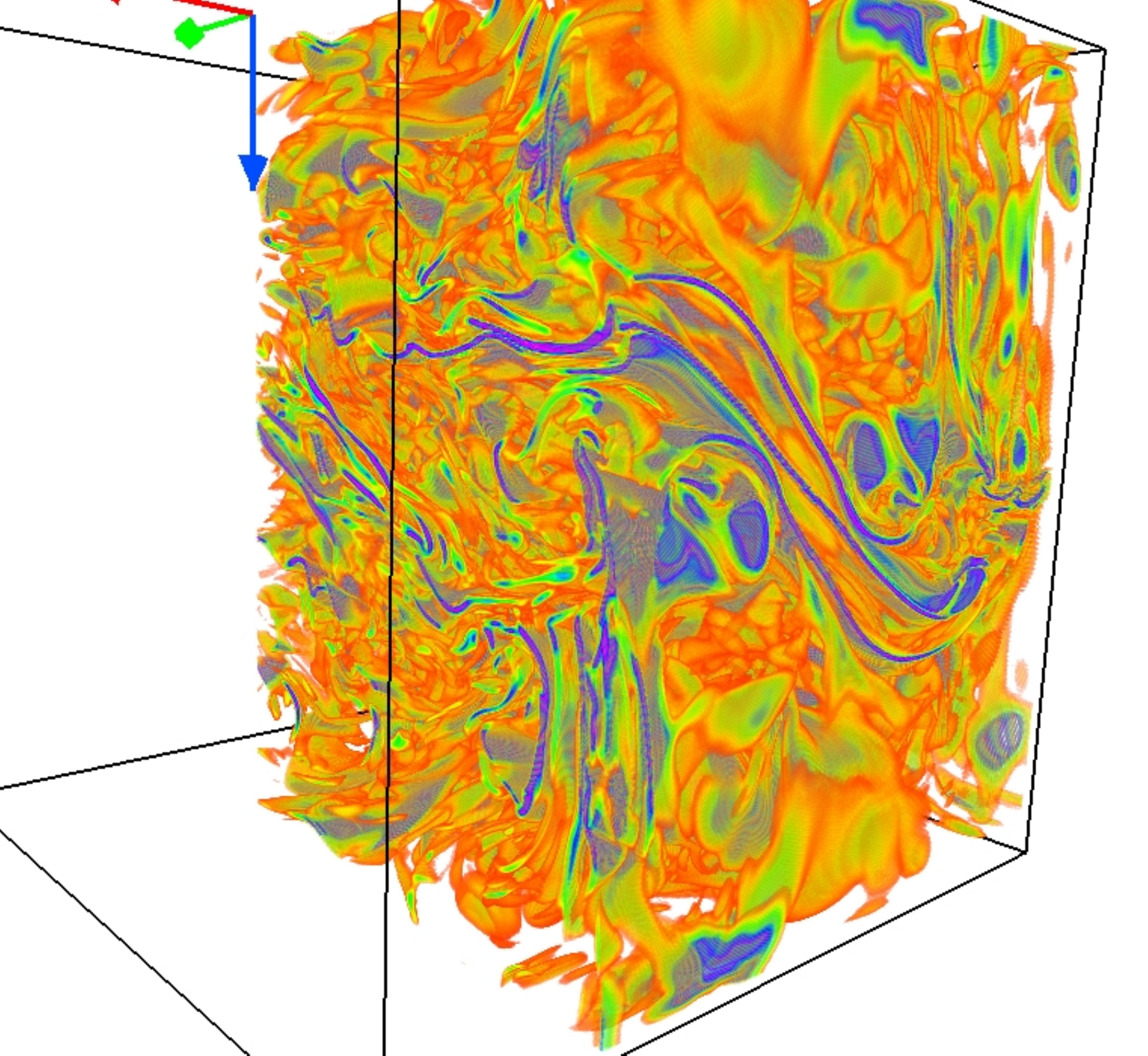}
\includegraphics[width=0.75\columnwidth]{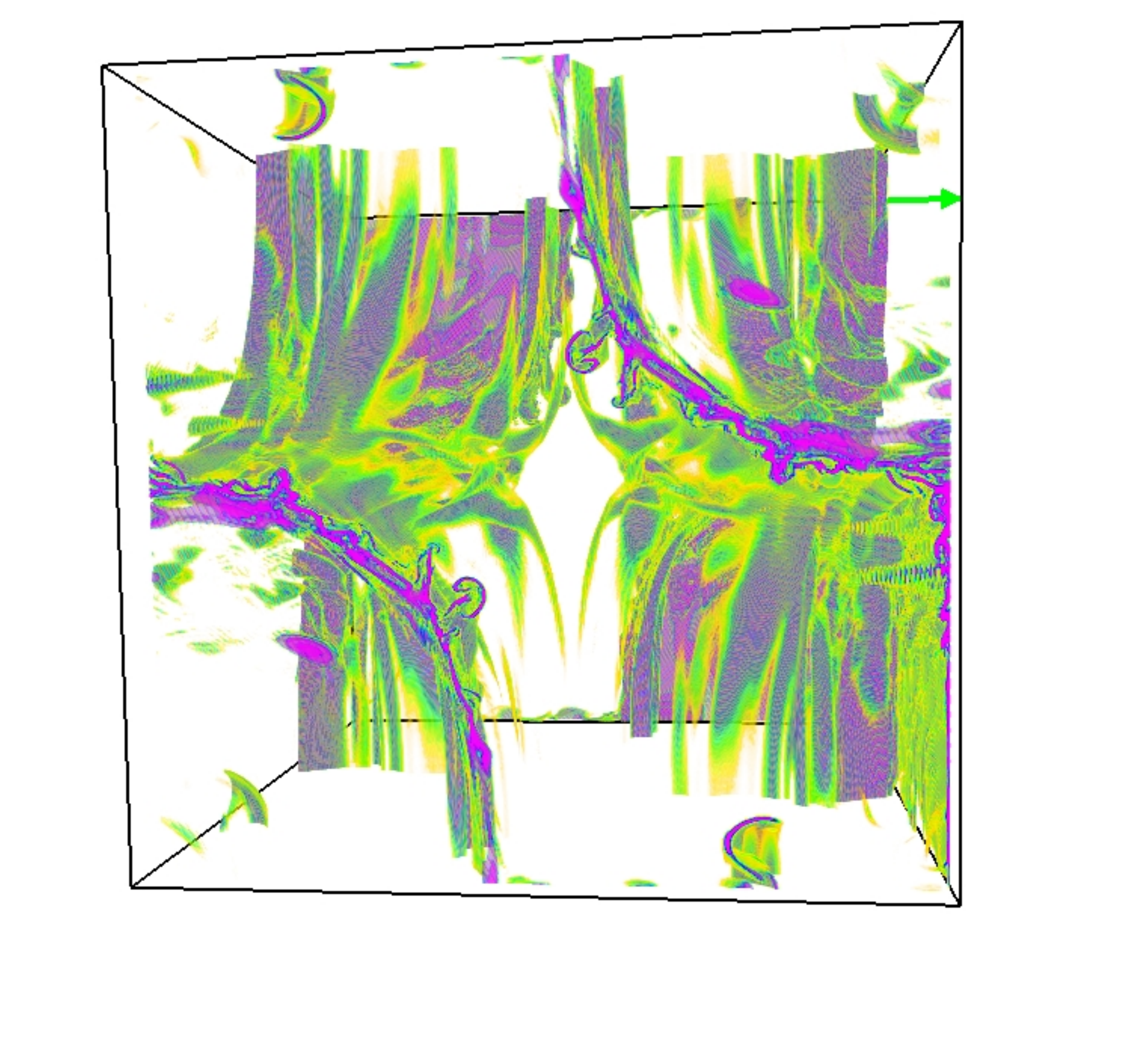}
\includegraphics[width=0.5\columnwidth]{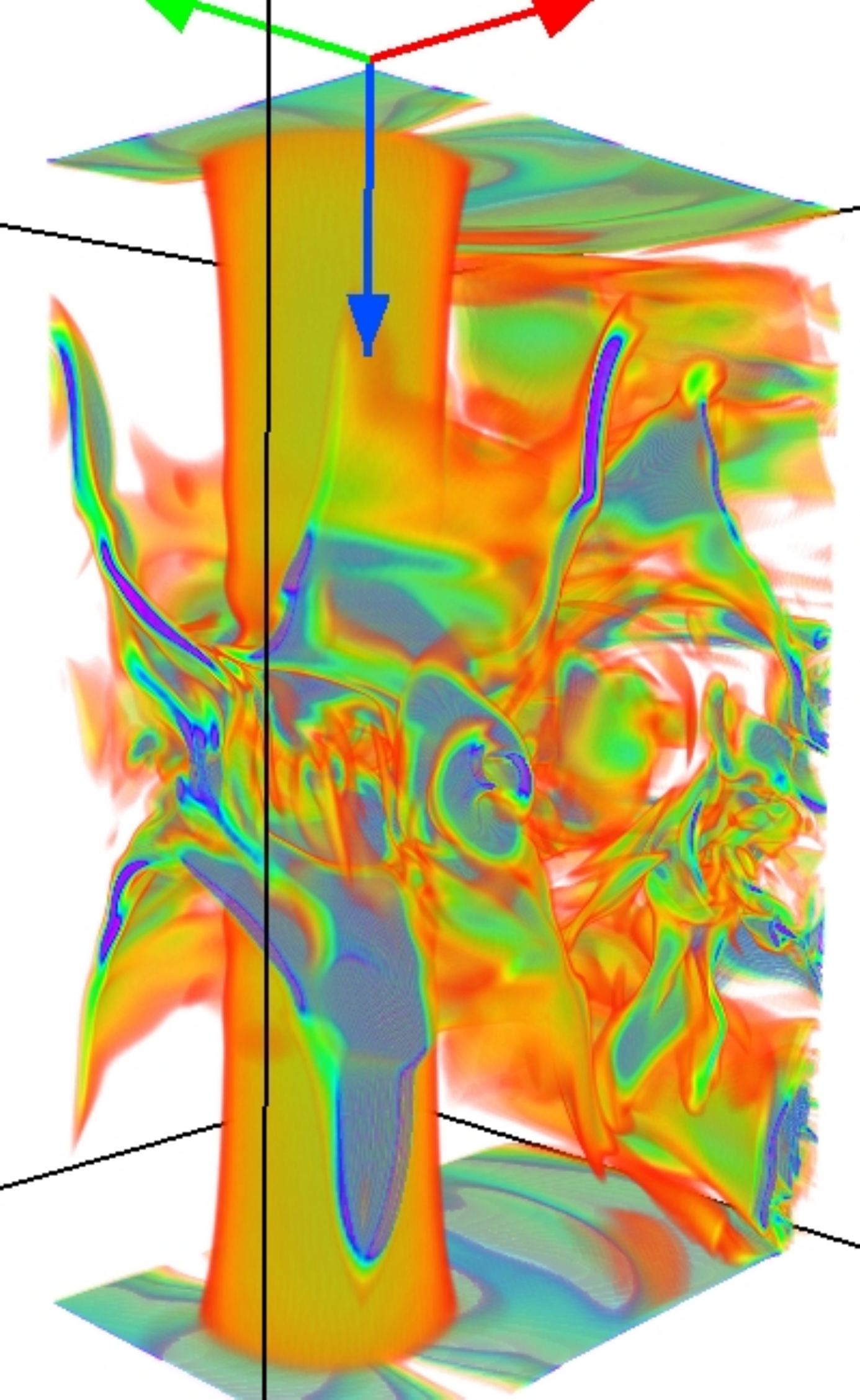}
\caption{(Color online) Visualizations of the magnetic energy at the end of the runs. Left: Run C2.
Center: Run I. Right: Run A.}
 \label{Fig:Viz} \end{figure*}

For completeness and in order to illustrate the physical-space distribution of magnetic energies in the the respective runs, we have performed some visualizations using the VAPOR software \cite{vapor}.
It is apparent on Fig.\ref{Fig:Viz} that large-scale coherent structures are present in the I and A flow. In contrast, the C-flow 
seems more isotropic. 
As discussed above it is a possibility that  steep structures, such as sharp and isolated current and vorticity sheets, can also lead to scaling in energy spectrum as it was reported in \cite{Alexakis_Dallas_Km2}.
This certainly looks as a possibility in the cases of I and A flows. Further studies of these structures are left for future work.

\section{Conclusions}\label{s:conclu}

In this paper, we have extended the analysis of non-universality of MHD spectra from the decaying case performed in \cite{lee2} to the forced case. We confirm the previous results, with either IK, K41 or WT spectra emerging on average, when using Taylor-Green forcing, including in the induction equation,
although temporal variations may be occurring.
Note that lack of universality in MHD has already been found by other authors, in the context of heating the solar corona \cite{dmitruk_03, mueller_05, rapazzo_07}, or in the presence of strong correlations between the velocity and the magnetic field \cite{grappin_83, politano_89, galtier_00}. We also confirm that these different scalings are linked to the magnetic energy content in the gravest mode and that, at least for these flows in which the four-fold symmetries of the TG vortex are imposed at all time, quasi-equipartition between the kinetic and magnetic energy obtains in the inertial range, with a variation of the ratio of the turn-over time to the Alfv\'en time consistent with the inertial index of the energy spectrum, as in the decaying case and contrary to the hypothesis made in \cite{GS}.
The influence of strong localized structures such as quasi-discontinuities can also alter energy scaling \cite{Alexakis_Dallas_Km2}.

 It should be noted that, even though there is no imposed magnetic field in these computations and thus no imposed anisotropy, an extension of this work could be to analyze the data in terms of anisotropic scaling with respect to a locally-defined quasi-uniform field, averaging the induction in a sphere of diameter the integral scale, as done for example in \cite{1536b} (see \cite{grappin_10} and references therein for anisotropic scaling in MHD). It is also well known that, in the atmosphere and the oceans, different spectra may emerge according to the relative strength of the stratification, the rotation and the forcing, as for example the Garrett-Munk versus Phillips spectra  \cite{ibra}.

There may be periods of evolution when the flow tends to one regime, and at other times to another regime. It was already observed in \cite{lee2} that for late times, the distinction between the K41 and IK regimes became difficult to make but of course the Reynolds number by then had decreased substantially. It is known that there are long-time fluctuations in most turbulent flows (\cite{dmitruk_11} and references therein) and this could lead to alternate exchanges of behavior, as already observed in \cite{gomez} in two-dimensional MHD, with turbulent periods of the order of one hundred turn-over times. These long-time fluctuations could lead to long-time fluctuations in spectral indices as well. Such behavior can be attributed to the lasting effects of nonlocal interactions between widely separated scales \cite{shell_I, shell_II} as observed in high-resolutions DNS of MHD turbulence.

It is not known whether these results will stand out in the limit of infinite Reynolds numbers, and higher Reynolds number computations will have to be performed in order to confirm the results presented in this paper.
There are other venues that can be taken as well: it is well-known that magnetic helicity and cross-helicity play essential roles in the dynamics oh MHD turbulence, and yet they are quenched by the symmetries in the present approach where symmetries are enforced at all times. It is already documented that, for the case of long time dynamics and perturbing the three initial conditions studied in this paper, vastly different regimes can be reached, with in particular the ratio of kinetic to magnetic energy varying in a large range \cite{julia}. This behavior can be understood in terms of minimization of energy subjected to the constraints of the invariance of $H_C$ and $H_M$ \cite{stribling_90, stribling_91}. This points out  to the possibly essential role played by the imposed four-fold symmetries. Such symmetries could be broken in part, as performed in \cite{giorgio_11}, leading to different modes growing in the dynamo regime.
A similar approach could be taken for the present problem of lack of universality, still allowing for some savings in computer resources compared with full-fledged MHD computations which might otherwise have to be performed. These issues will need more investigations.

\begin{acknowledgments}  
Computations were performed at IDRIS, NCAR and at  M\'esocentre SIGAMM hosted at the Observatoire de la C\^ote d'Azur, where the visualizations were also done.
The National Center for Atmospheric Research is sponsored by the National Science Foundation.  
Marc  Brachet acknowledges a GTP fund allocation. 
\end{acknowledgments}


\begin{thebibliography}{23}
\bibitem{vapor}
VAPOR is a freeware developed at the National Center for Atmospheric Research, see ww.vapor.ucar.edu.
\end{thebibliography}
\end{document}